# Disentangle magnon magnetoresistance from anisotropic and spin Hall magnetoresistance in NiFe/Pt bilayers

Yanjun Xu, Yumeng Yang, Ziyan Luo and Yihong Wu*

*Department of Electrical and Computer Engineering, National University of Singapore, 4 Engineering Drive 3, Singapore 117583, Singapore*

We conducted a systematic angular dependence study of nonlinear magnetoresistance in NiFe/Pt bilayers at variable temperatures and fields using the Wheatstone bridge method. We successfully disentangled magnon magnetoresistance from other types of magnetoresistances based on their different temperature and field dependences. Both the spin Hall / anisotropic and magnon magnetoresistances contain $\sin\varphi_m$ and $\sin 3\varphi_m$ components ($\varphi_m$: angle between current and magnetization), but they exhibit different field and temperature dependences. The competition between different types of magnetoresistances leads to a sign reversal of $\sin 3\varphi_m$ component at a specific magnetic field, which was not reported previously. The phenomenological model developed is able to account for the experimental results for both the NiFe/Pt and NiFe/Ta samples with different thicknesses of the constituent layers. Our results demonstrate the importance of disentangling different types of magnetoresistances when characterizing the charge-spin interconversion process in magnetic heterostructures.

*Author to whom correspondence should be addressed: elewuyh@nus.edu.sg

# I. INTRODUCTION

Spin-charge interconversion in ferromagnet (FM)/heavy metal (HM) bilayers has been a subject of intensive studies [1,2]. When a charge current passes through a FM/HM bilayers, spin accumulation occurs at the interface due to either bulk spin Hall effect (SHE) [3-6] of the HM or Rashba-Edelstein effect (REE) at the interface [7-9], or the combination of both effects. If SHE is dominant as is the case for many FM/HM systems, the non-equilibrium spins affect both the magnetization dynamics of the FM layer and spin-dependent carrier transport of the bilayers in several aspects [1,2]. First, spins with their polarization misaligned with the magnetization direction of the FM layer are absorbed by it, exerting spin-orbit torques (SOT) [1,2,10,11] on the magnetization. The SOT, which has been observed in a variety of FM/HM bilayer systems, provides an efficient way to manipulate the magnetization of ultrathin FM layers [1,2,10-12]. Second, spins with the polarization parallel to the magnetization direction are partially reflected/transmitted at the FM/HM interface; the reflected spins flow back to the HM layer, inducing an additional resistance called spin Hall magnetoresistance (SMR) [13,14]. For FM with an in-plane magnetic anisotropy, the SMR exhibits the same angular dependence as that of anisotropic magnetoresistance (AMR), *i.e.*, proportional to $cos^2\varphi_m$, where $\varphi_m$ is the angle between the magnetization and current direction. Although it is still a subject of debate [15], it is commonly believed that the SMR is originated from the inverse spin Hall effect (ISHE) of HM [13,16]. Third, the interfacial/transmitted non-equilibrium spins interact with the FM layer, giving rise to a magnetoresistance which is odd under either magnetization or current reversal, and is called unidirectional spin Hall magnetoresistance (USMR) [17].

The USMR at saturation state is interpreted as originated from spin-dependent electron scattering both at the FM/HM interface and inside the FM layer, and is independent of the external field strength (hereafter we refer it to as SD-USMR) [17]. But, recent studies at low-fields unveil a rather complex scenario due to the excitation of magnons in the FM layer, particularly at high excitation current [18-22]. The magnon-induced magnetoresistance was found to be asymmetrical with respect to both the external field ($H_{ex}$) and current density ($j$), and scales as $H_{ex}^{-p}$ and $j + j^3$, respectively [20]. The exponent *p*, with the value close to 1, is observed to increase with current and decrease with the FM layer thickness and is believed to be related to the stiffness of magnon modes [20]. In addition to the field strength, the magnon magnetoresistance (MMR) is also found to contain odd harmonics of $\varphi_m$ [20]. Although the experimental data can be accounted for reasonably well using phenomenological models, the origin of both the $H_{ex}$ and $\varphi_m$ – dependence is still not well understood. The difficulty lies in the fact that the 2nd harmonic technique that is commonly used to measure the non-linear MR is unable to distinguish the current induced MR contributions from different sources. Therefore, in order to gain an insight of spin-charge interconversion in FM/HM layers, it is important to disentangle the contributions from different sources and examine how they contribute specifically to the overall MR signal measured by the 2nd harmonic technique. In this context, we have conducted a systematic angular dependence study of MR in NiFe/Pt bilayers at variable temperature, field and current density, which allows us to disentangle the MMR from AMR and SMR by using a newly developed bridge technique. We develop an analytical model in which we argue that, instead of a power-law dependence on external field ($H_{ex}$) with exponent *p*, all the MMR terms scale as $(H_{ex} + H_m)^{-1}$, where $H_m$ is an internal induction field proportional to the saturation magnetization. We show that both the MMR and 2nd order AMR/SMR contain $\sin \varphi_m$ and $\sin 3\varphi_m$ dependence on $\varphi_m$, but the sign of the $\sin 3\varphi_m$ term is opposite for MMR and AMR/SMR. For a typical NiFe/Pt bilayers, the MMR is 2 to 3 orders smaller than the first order AMR/SMR, but it is comparable to the magnitude of 2nd order AMR/SMR, and therefore, the competition between MMR and AMR/SMR can lead to sign change of the $\sin 3\varphi_m$ component at a specific magnetic field. The analysis based on the proposed model corroborates well with the experimental results. Our findings shed light on the roles of magnons in carrier transport of FM/HM bilayers and demonstrate the importance of disentangling different types of MRs when characterizing charge-spin interconversion and related effects in FM/HM bilayers.

## II. EXPERIMENTAL DETAILS

The samples used for the current-induced MR measurements are NiFe($t_{NiFe}$)/Pt($t_{Pt}$) bilayers deposited on SiO$_2$/Si substrates, unless otherwise specified. All the layers were prepared by dc magnetron sputtering with a base pressure of $2 \times 10^{-8}$ Torr and working pressure of $3 \times 10^{-3}$ Torr, respectively. Here, the numbers inside the brackets denote layer thickness in nm. Instead of using the standard lock-in technique, in this work we employ the Wheatstone bridge method to measure the 2$^{nd}$ harmonic signal induced by an ac current [23-25]. Figure 1(a) shows the scanning electron micrograph (SEM) of the fabricated Wheatstone bridge sample comprising of four identical ellipsoidal NiFe/Pt elements with a long axis length (a) of 800 μm and an aspect ratio of 4:1. The spacing (L) between the two electrodes for each element is a/3. During the measurement, the ac current is applied to the top and bottom terminals by Keithley 6221 current source without grounding. The Keithley 2182A nano-voltmeter is used to capture the time average or dc output voltage from the left and right bridge terminals. All the measurements were performed inside a Quantum Design Versalab Physical Property Measurement System with a temperature range of 50 – 400K and a field range of 0 – 3T.

## III. DISENTANGLE MMR FROM AMR AND SMR by BRIDGE TECHNIQUE

Before discussing the experimental results, we first explain how an ac driven bridge technique can be used to disentangle MMR from AMR and SMR in FM/HM bilayers [23-26]. As shown in the schematic of Fig. 1(b), the samples in this work are configured in a Wheatstone bridge structure which comprises of four ellipsoidal NiFe/Pt elements. When a current is applied to the bridge, it flows along the long axis of the elements, but the direction is opposite for element 1 and 4 versus 2 and 3. This means that the spin polarization ($\sigma$) induced by the current in the two pairs of elements, *i.e.*, 1 and 4 versus 2 and 3, is also opposite to each other. This facilitates measurement of 2$^{nd}$ order AMR/SMR because the current induced resistance change in the adjacent elements also exhibits opposite sign. In addition to AMR/SMR, the bridge is also a natural method to measure SD-USMR and MMR because the SD-USMR exhibits a $\sin\varphi_m$ dependence, whereas the MMR depends on both $\sin\varphi_m$ and $\sin 3\varphi_m$ with $\varphi_m$ the angle between current and magnetization. The $\sin\varphi_m$ dependence of SD-USMR is explained in literature based on the analogy with giant magnetoresistance wherein the interface is assumed to play the role of one of the FM layers[17,20]. However, the origin of $\sin 3\varphi_m$ term remains unexplained so far. As elaborated below, we argue that the $\sin 3\varphi_m$ term is due to the combined effect of angel-dependent magnon-excitation and the AMR. The magnon excitation efficiency in FM/HM bilayers is proportional to $\sin\varphi_m$ because it is strongest when $\boldsymbol{\sigma} \parallel \boldsymbol{M}$. The excited magnon increases the resistance via electron-magnon scattering which, in the 1$^{st}$ order, should also exhibit a $\sin\varphi_m$ dependence. As the scattering involves spin-flip, it will naturally affect the AMR since the latter is resulted from mixing of spin-up and spin-down electrons via spin-orbit interaction [27]. Therefore, we may conjecture that, in addition to the $\sin\varphi_m$ term, there is another MMR related term which is proportional to $\sin\varphi_m \cos^2\varphi_m$, or alternatively, it may be written in the form of $(\sin\varphi_m + \sin 3\varphi_m)/4$. According to Fert and Campbell [27], the AMR may be considered as being cause by "resistivity transfer" from spin-down electrons to spin-up electrons, and the transfer is strongest when the magnetization is parallel to current. Therefore, the sign of $\sin\varphi_m \cos^2\varphi_m$ term should be opposite to that of the $\sin\varphi_m$ term. The two terms combined give an MMR in the form of $(\Delta R_{n1}\sin\varphi_m + \Delta R_{n3}\sin 3\varphi_m)$. The absolute sign of the $\sin\varphi_m$ and $\sin\varphi_m \cos^2\varphi_m$ terms depends on the deposition sequence of the FM/HM bilayers and the sign of HM Hall angle. The two terms combined determine the sign of $\Delta R_{n1}$ and $\Delta R_{n3}$.

By including the contributions from AMR/SMR, SD-USMR, MMR, and thermoelectric effect [17,20], the longitudinal resistance element 1 ($R_1$) in the bridge may be expressed as

$$R_1 = R_0 + \Delta R_{AMR} \cos^2(\varphi_m + \Delta\varphi_m) + \Delta R_{n1} \sin(\varphi_m + \Delta\varphi_m) + \Delta R_{n3} \sin 3(\varphi_m + \Delta\varphi_m) - \Delta R_0 \sin(\varphi_m + \Delta\varphi_m) \qquad (1)$$

where $R_0$ is the longitudinal resistance when the magnetization is perpendicular to the current direction, $\Delta R_{AMR}$ contains the MR contributions from both AMR and SMR, $\varphi_m$ is the angle between current and magnetization at zero current (set by the external field), $\Delta\varphi_m$ is the angle change induced by the current, $\Delta R_{n1}$ and $\Delta R_{n3}$ represent $sin\varphi_m$ and $sin3\varphi_m$ contribution to the MMR, respectively, and $\Delta R_0$ is the field-independent MR term, which is proportional to the current and accounts for both the SD-USMR and thermoelectric effect [17]. $\Delta R_{n1}$ and $\Delta R_{n3}$ are dependent on both the current and external field. For the in-plane magnetized NiFe/Pt structure, $\Delta\varphi_m$ is approximately given by $\frac{h_{FL}^{SH}\cos\varphi_m}{H_{ex}}$ with the SOT induced field-like effective field $h_{FL}^{SH} = \alpha_{FL}^{SH} j_{Pt}$, where $\alpha_{FL}^{SH}$ is the SOT efficiency, $H_{ex}$ is the external field, and $j_{Pt}$ is the current density in Pt layer.

Similarly, the longitudinal resistance element 2 ($R_2$) can be expressed as (note : $\Delta\varphi_m$ has a negative sign with respect to element 1 and $\varphi_m$ differs from that of element 1 by $\pi$)

$$R_2 = R_0 + \Delta R_{AMR} \cos^2(\varphi_m - \Delta\varphi_m) - \Delta R_{n1} \sin(\varphi_m - \Delta\varphi_m) - \Delta R_{n3} \sin 3(\varphi_m - \Delta\varphi_m) + \Delta R_0 \sin(\varphi_m - \Delta\varphi_m). \qquad (2)$$

The sign of last three terms is opposite to that of Eq. (1) because of reversal of current direction in element 2 against element 1. When an ac current $I = I_0 \sin\omega t$ is applied to the bridge, the output voltage of the Wheatstone bridge is given by $V_b = V_2 - V_1 = \frac{I_0 \sin\omega t}{2}(R_2 - R_1)$. We used an ac current instead of dc current because it suppresses the thermal drift and reduces noise [24,25]. Since the last three terms in both Eqs. (1) and (2) are proportional to the current density at low current region, we may write them as $\Delta R_{n1} = \Delta r_{n1} j_{Pt}$, $\Delta R_{n3} = \Delta r_{n3} j_{Pt}$, and $\Delta R_0 = \Delta r_0 j_{Pt}$, where $\Delta r_{n1}$, $\Delta r_{n3}$, and $\Delta r_0$ are current-independent coefficients. The 2$^{nd}$ term is implicitly dependent on the current through the SOT-induced change in $\varphi_m$. After some algebra (see details in Supplementary Material), we can obtain $V_b = \frac{\Delta R I_0}{2} - \frac{\Delta R I_0}{2} \cos 2\omega t$, where:

$$\Delta R = \Delta R_{AMR} \frac{\alpha_{FL}^{SH} j_{Pt0}}{H_{ex}} \left( \frac{\sin\varphi_m + \sin 3\varphi_m}{2} \right) - \Delta r_{n1} j_{Pt0} \sin\varphi_m - \Delta r_{n3} j_{Pt0} \sin 3\varphi_m$$

$$+ \Delta r_0 j_{Pt0} \sin\varphi_m \qquad (3)$$

where $j_{Pt0}$ is the amplitude of the current density in the Pt layer. The time average or dc output voltage is $\Delta R I_0/2$, which can be directly measured by the dc nano-voltmeter. As the dc output is proportional to $\Delta R$, it can be used to characterize the spin-charge interconversion process without resorting to the lock-in technique. In fact, the bridge technique is generic, and can be used to characterize any type of current-induced nonlinear resistances as long as it is an odd function of current. Eq.(3) can be further reduced in the form of $\Delta R(\varphi) = \Delta R_{\varphi_m} \sin\varphi_m + \Delta R_{3\varphi_m} \sin 3\varphi_m$ with $\Delta R_{\varphi_m} = -\Delta r_{n1} j_{Pt0} + \frac{1}{2} \frac{\Delta R_{AMR} \alpha_{FL}^{SH} j_{Pt0}}{H_{ex}} + \Delta r_0 j_{Pt0}$ and $\Delta R_{3\varphi_m} = -\Delta r_{n3} j_{Pt0} + \frac{1}{2} \frac{\Delta R_{AMR} \alpha_{FL}^{SH} j_{Pt0}}{H_{ex}}$. Considering the fact that, for bulk materials, the MMR scales with the external field as $\frac{1}{H_{ex}+H_m}$, where $H_m$ is an induced internal field proportional to the magnetization [28-30], we may write $\Delta R_{\varphi_m}$ and $\Delta R_{3\varphi_m}$ as:

$$\Delta R_{\varphi_m} = \frac{A}{H_{ex}} + \frac{B}{H_{ex}+H_m} + \Delta r_0 j_{Pt0} \qquad (4)$$

$$\Delta R_{3\varphi_m} = \frac{A}{H_{ex}} + \frac{C}{H_{ex}+H_m} \qquad (5)$$

where $A = \frac{1}{2}\Delta R_{AMR}\alpha_{FL}^{SH}j_{Pt0}$, $B = \alpha_m j_{Pt0}$, and $C = \beta_m j_{Pt0}$. $\alpha_m$ and $\beta_m$ are field-independent constants, which are related to $\Delta r_{n1}$ and $\Delta r_{n3}$ by $\alpha_m = -\Delta r_{n1}(H_{ex} + H_m)$ and $\beta_m = -\Delta r_{n3}(H_{ex} + H_m)$, respectively. Eqs. (4) and (5) are the central equations which will be used to discuss the experimental results.

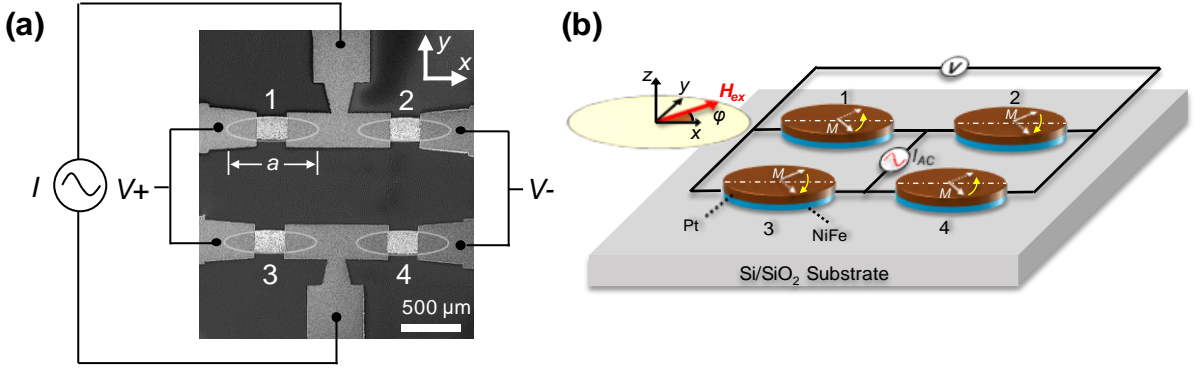

FIG. 1. (a) Scanning electron micrograph of NiFe/Pt bilayer Wheatstone bridge. (b) Schematic of the Wheatstone bridge comprised of four ellipsoidal NiFe/Pt bilayer elements with the arrows indicating the magnetization direction. The external field is rotated in *xy* plane with angle $\varphi$ relative to the current direction. Scale bar: 500 μm.

## IV. RESULTS AND DISCUSSION

### A. External field dependence of nonlinear magnetoresistance

We first present the results of angle dependent nonlinear MR in an ac excited NiFe(1.8)/Pt(2) Wheatstone bridge measured at room temperature. Figure 2(a) shows the field angle dependence of the bridge output voltage $V_{out}$ (hereafter we refer it to as the dc output) at different field strengths $H_{ex}$ = 110, 140, 170, 200 and 500 Oe. Results for a wider field range from 10 Oe to 2 T are given in the Supplementary Material. During the measurements, an ac current with a root mean square (RMS) amplitude density of $5.5 \times 10^5$ A/cm² (in the Pt layer) and frequency of 5000 Hz was applied to the bridge and its output voltage $V_{out}$ was recorded using a nano-voltmeter. The current density in the Pt layer is calculated based on the experimentally extracted resistivities of the Pt and NiFe layers. It is worth emphasizing that all the curves in the figure are directly plotted from the raw data without any averaging or offset compensation. The surprisingly low noise level demonstrates the superior quality of the signal obtained by the bridge method. The general observations can be summarized as follows. When $H_{ex}$ is small (< 50 Oe), the $V_{out}(\varphi)$ curves exhibit complex shapes as the field is insufficient to align the magnetization into the external field direction due to the anisotropic field. The curves begin to show a more consistent pattern when the field exceeds 50 Oe, *i.e.*, when $\varphi_m \approx \varphi$, where $\varphi_m$ is the azimuth angle of magnetization. In the intermediate range, from 50 Oe to 10 kOe, the $V_{out}(\varphi)$ curves can be decomposed into two components with distinct $\varphi$ dependence, *i.e.*, $V_{out}(\varphi) = V_\varphi \sin \varphi + V_{3\varphi} \sin 3\varphi$, as demonstrated by the fitting curves in Fig. 2(a) (solid-lines). When $H_{ex}$ increases further, the $\sin 3\varphi$ component eventually becomes diminishingly small, leaving only the $\sin \varphi$ component. We now turn to the $\sin 3\varphi$ component in the intermediate field range. Figure 2(b) shows the $\sin 3\varphi$ component decomposed from the $V_{out}(\varphi)$ curves shown in Fig. 2(a) by subtracting out the $\sin \varphi$ component. It is interesting to note that the amplitude of the $\sin 3\varphi$ component initially decreases with increasing the external field, approaches zero at $H_{ex} \approx 170 \, Oe$, after which the sign reverses but the amplitude increases again up to $H_{ex} \approx 500 \, Oe$. Beyond this field, it decreases monotonically and becomes almost undetectable at $H_{ex} \approx 20000 \, Oe$. This kind of behaviour can be readily understood by two field dependent MR terms with opposite polarity and different external field dependences.

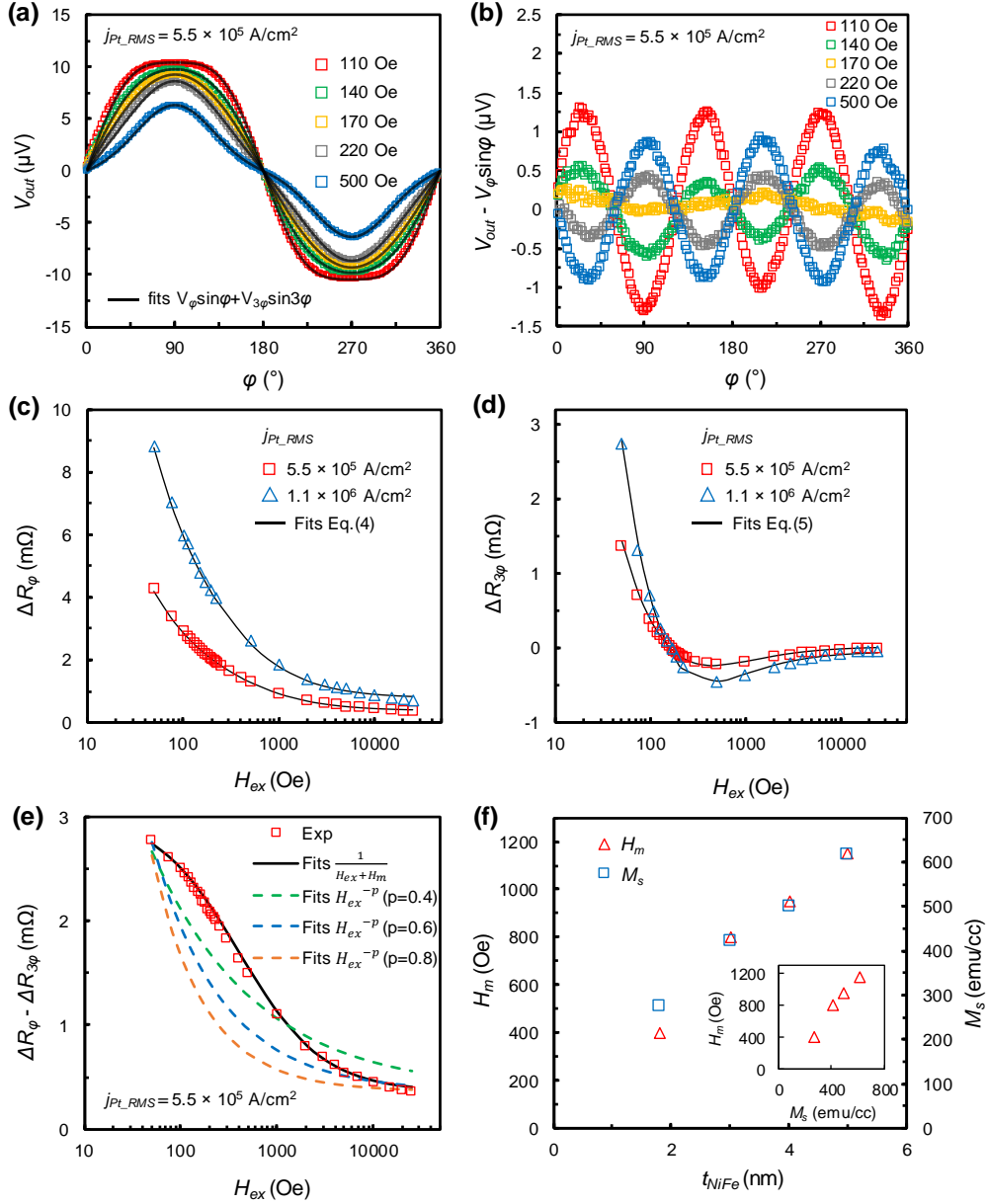

FIG. 2. (a) Bridge output voltage as a function of the in-plane field angle $\varphi$ at different field strength: 110, 140, 170, 220 and 500 Oe together with the fitting curves (solid-liness). The rms amplitude of the current density in Pt layer is $5.5 \times 10^5$ A/cm$^2$ and frequency of the driving current is 5000 Hz. (b) The $\sin 3\varphi$ component decomposed from the $V_{out}(\varphi)$ curves in (a) by subtracting out the $\sin \varphi$ component. (c,d) $\Delta R_\varphi$ (c) and $\Delta R_{3\varphi}$ (d) as a function of $H_{ex}$ for two different current densities (square and triangle). The solid-lines are fitting results using Eq. (4) (for $\Delta R_\varphi$) and Eq. (5) (for $\Delta R_{3\varphi}$), respectively. (e) The extracted values of $\Delta R_\varphi - \Delta R_{3\varphi}$ as a function of $H_{ex}$ (square) together with the fitting curves based on different external field dependence. (f) Comparison of experimentally extracted $H_m$ and saturation magnetization at different NiFe thickness in NiFe($t_{NiFe}$)/Pt(2) bilayer structures. Inset: dependence of $H_m$ on $M_s$.

To have a more quantitative understanding of the field dependence of the nonlinear MR signal, we extract $V_\varphi$ and $V_{3\varphi}$ and then divide by $I_0/2$ to obtain nonlinear resistance components $\Delta R_\varphi$ and $\Delta R_{3\varphi}$. The results are shown in Fig. 2(c) and Fig. 2(d), respectively, as a function of $H_{ex}$ for two different current densities (square: $5.5 \times 10^5 \text{A/cm}^2$ and triangle: $1.1\times 10^6 \text{A/cm}^2$). The solid-lines are fitting results using the Eqs. (4) and (5). The parameters used for the fittings are $A = 130 \text{ m}\Omega \cdot \text{Oe}$, $B = 637 \text{ m}\Omega \cdot \text{Oe}$, $C = -433 \text{ m}\Omega \cdot \text{Oe}$, $H_m = 400 \text{ Oe}$, and $\Delta r_0 j_{Pt0} = 0.36 \text{ m}\Omega$ for $j_{Pt\_RMS} = 5.5 \times 10^5 \text{ A/cm}^2$, and $A = 256.0 \text{ m}\Omega \cdot \text{Oe}$, $B = 1300 \text{ m}\Omega \cdot \text{Oe}$, $C = -850 \text{ m}\Omega \cdot \text{Oe}$, $H_m = 400 \text{ Oe}$, and $\Delta r_0 j_{Pt0} = 0.72 \text{ m}\Omega$ for $j_{Pt\_RMS} = 1.1 \times 10^6 \text{ A/cm}^2$, where $j_{Pt\_RMS}$ is the rms amplitude of the current density in the Pt layer. The fitting is surprisingly good for both $\Delta R_\varphi$ and $\Delta R_{3\varphi}$ in the entire field range. In addition to the fast decay at small field, the opposite sign of the first and second terms in Eq. (5) well explains the sign reversal for $\Delta R_{3\varphi}$ as the external field increases.

In order to compare with the model proposed previously by Avci *et al.* [20], we have also tried to fit the field dependence of magnon contribution to the MR using the power law, *i.e.*, $H_{ex}^{-p}$. According to Eqs. (4) and (5), the field dependence of MMR is directly reflected in the difference between $\Delta R_\varphi$ and $\Delta R_{3\varphi}$ because $\Delta R_\varphi - \Delta R_{3\varphi} = \frac{B-C}{H_{ex}+H_m} + \Delta r_0 j_{Pt0}$ in which the 2nd term is independent of the field. Fig. 2(e) shows the fitting of $\Delta R_\varphi - \Delta R_{3\varphi}$ versus $H_{ex}$ for the NiFe(1.8)/Pt(2) sample at $5.5 \times 10^5 \text{A/cm}^2$. As expected, the experimental results (square symbol) can be fitted well using the present model (solid-lines) with $B = 637 \text{ m}\Omega \cdot \text{Oe}$, $C = -433 \text{ m}\Omega \cdot \text{Oe}$, $H_m = 400 \text{ Oe}$ and $\Delta r_0 j_{Pt0} = 0.36 \text{ m}\Omega$. In contrast, the fitting curves using $aH_{ex}^{-p} + \Delta r_0 j_{Pt0}$ (where $a$ is a fitting constant) can hardly fit the experimental results even by varying $p$ in a large range as shown in the dotted-lines. This result suggests that the $1/(H_{ex}+H_m)$ dependence is more appropriate to describe the effect of external field on MMR. As we mentioned earlier, the $H_m$ is the induced internal field which, has been shown previously to be proportional to the saturation magnetization ($M_s$) [28-30]. To further investigate the origin of $H_m$, we have fabricated a series of NiFe($t_{NiFe}$)/Pt(2) samples with different NiFe thicknesses. We then repeated the measurements to extract $H_m$ for samples with different NiFe thicknesses and the results are plotted in Fig. 2(f). Also shown in the figure is the $M_s$ at different NiFe thicknesses. We can see that both $H_m$ and $M_s$ scale almost linearly with the thickness except for the sample with at a NiFe thickness of 2 nm. The inset shows the plot of $H_m$ v.s. $M_s$. A nearly linear relationship between $H_m$ and $M_s$ is clearly seen, which further ascertains the validity of the present model.

### B. Current density dependence of nonlinear magnetoresistance

After successfully decomposing the nonlinear MR into components of different origins, we now examine its dependence on current density. Fig. 3(a) and Fig. 3(b) show the current density dependence of $\Delta R_\varphi$ and $\Delta R_{3\varphi}$ at different external fields (symbols are experimental results and solid-lines are linear fittings). A linear-dependence on the rms current density is obtained for both $\Delta R_\varphi$ and $\Delta R_{3\varphi}$ at all fields in the low current density range of $1.8 \times 10^5 \sim 1.1 \times 10^6$ A/cm$^2$. Interestingly, although the slope of the fitting lines for $\Delta R_{3\varphi}$ changes sign from positive to negative, the linear relationship remains in the entire range, which suggest that all the nonlinear resistance terms are induced by the current. This can be well explained by Eq. (3) in which all terms with angle dependence of $sin\varphi_m$ and $sin3\varphi_m$ are proportional to the current density. However, when the current density exceeds $2 \times 10^6$ A/cm$^2$, both $\Delta R_\varphi$ and $\Delta R_{3\varphi}$ start to exhibit a nonlinear dependence on current at low field, but the linear dependence restores at high field. Fig. 3c and Fig.3d show the typical results at low (1000 Oe) and high (20000 Oe) field, respectively. Similar results have been obtained previously for Pt/Co bilayers in which the rapid increase of nonlinear MR at large current density is attributed to the contribution from thermal magnons [20,21]. The dependence on $\varphi_m$ remains the same for large current suggests that the thermal magnon excitation in FM/HM bilayers is correlated with the magnons stimulated by the SHE-generated non-equilibrium spins from the HM layer.

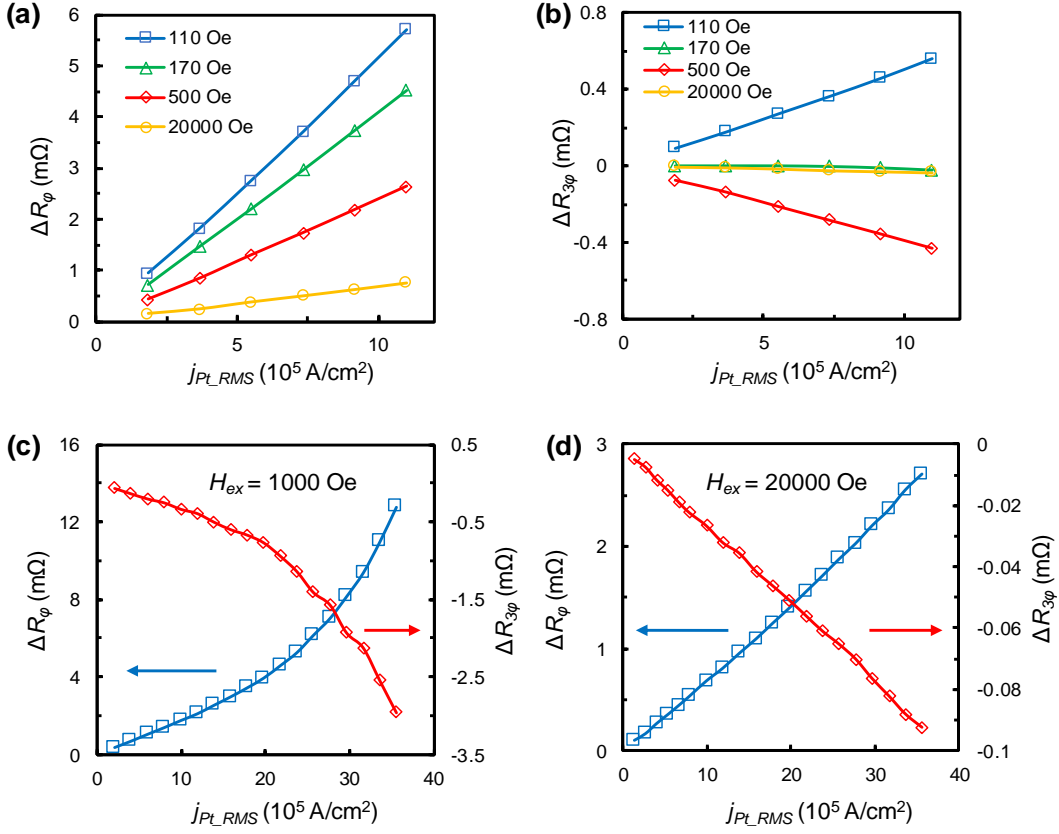

FIG. 3. Dependence of the nonlinear resistance on the current density. (a,b) Dependence of $\Delta R_\varphi$ (a) and $\Delta R_{3\varphi}$ (b) on current density in the range of $1.8 \times 10^5$ - $1.1 \times 10^6$ A/cm$^2$ at different applied field: 110, 170, 500 and 20000 Oe. (c,d) Dependence of $\Delta R_\varphi$ and $\Delta R_{3\varphi}$ in the large current density range and at an applied field of 1000 Oe (c) and 20000 Oe (d).

## C. Temperature dependence of nonlinear magnetoresistance

In order to further substantiate the argument which leads to Eqs. (4) and (5), we investigate the temperature dependence of the fitting parameters *A, B*, and *C* from 50 K to 300 K. The low temperature limit of 50 K is set by the measurement system used. The sample used was the same as the one whose room temperature characteristics have already been presented in Fig. 2. The Pt current density was fixed at $5.5 \times 10^5$ A/cm$^2$ unless otherwise specified.

Figures 4(a) and 4(b) show the field-dependence of $\Delta R_\varphi$ and $\Delta R_{3\varphi}$ at 300 K (square), 200 K (triangle) and 50 K (circle), respectively. Reasonably good fittings have been obtained for both $\Delta R_\varphi$ and $\Delta R_{3\varphi}$ at all temperatures (solid-lines) using Eqs. (4) and (5). The fitting parameters $A$, $B$ and $C$ as a function of temperature are summarized in Fig. 4(c). As can be seen, $A$ behaves quite differently from $B$ and $C$ with respect to the temperature. The value $A$ decreases slightly from 50 K to 300 K though there is a small kink at around 200 K above which it is almost constant. In contrast, $B$ and $C$ are very small at 50 K, and when the temperature increases, it initially increases slowly from 50 K to 200 K, but the rate of increase becomes much larger between 200 K – 300 K. This demonstrates clearly the different origins of $A$ versus $B$ and $C$. The fast increase of $B$ and $C$ with temperature further confirms that the $B$ and $C$ related terms in Eqs. (4) and (5) are due to magnon excitation, whereas $A$ is mainly from the combined effect of AMR and SMR.

To have a more quantitative view of the contributions from AMR/SMR and MMR, we have extracted values of $\Delta R_1$ and $\Delta R_3$ at 100 Oe, 500 Oe and 20000 Oe with $j_{Pt\_RMS} = 5.5 \times 10^5$ A/cm² and compare them with that of experimentally measured $\Delta R_{AMR}$ from 50 K to 300 K in Fig. 4(d). As can be seen, $\Delta R_1$ is negative and $\Delta R_3$ (solid-lines with symbols) is positive in the entire temperature and field range investigated. The absolute values of both MMR terms decrease with increasing either the external field or temperature. At 100 Oe and 300 K, both $\Delta R_1$ and $\Delta R_3$ are in the $m\Omega$ range, but they are diminishingly small at a field of 3 T. In contrast, $\Delta R_{AMR}$ decreases with increasing the temperature (dotted-line with diamond symbol). At 300 K, $\Delta R_1$ and $\Delta R_3$ are 2-3 orders smaller than $\Delta R_{AMR}$. This is understandable because $\Delta R_{AMR}$ is the 1st order, whereas $\Delta R_1$ and $\Delta R_3$ are 2nd order signals. On the other hand, the 2nd order AMR/SMR, *i.e.*, the first term of Eq. (3), is comparable to $\Delta R_1$ and $\Delta R_3$ as $\alpha_{FL}^{SH}$ is around 0.76 Oe/$[\frac{10^6 A}{cm^2}]$ for NiFe(1.8)/Pt(2) [23]. Our results demonstrate clearly that the MMR contribution is important even at low current density when charactering the spin-charge interconversion phenomena in FM/HM bilayers such as the spin-orbit torque. Depending on the nature of signal that is measured as well as the measurement condition, both the amplitude and sign can be different.

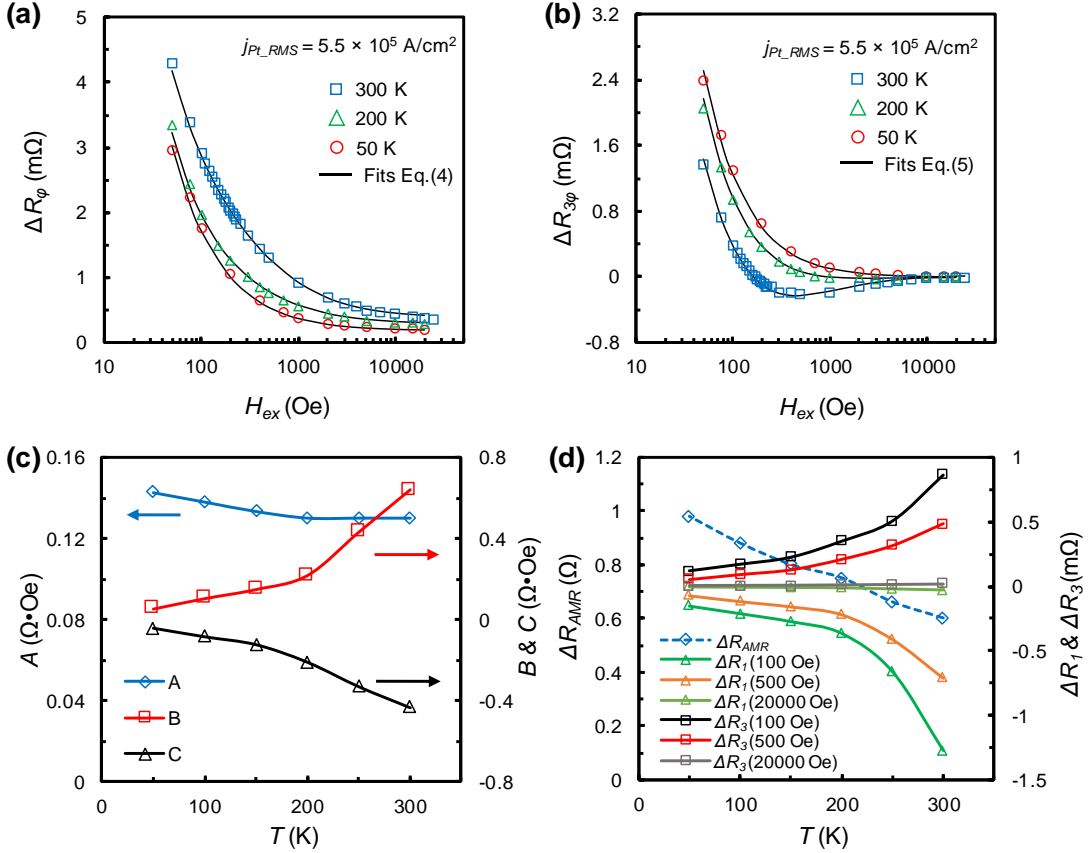

FIG. 4. (a, b) $\Delta R_\varphi$ (a) and $\Delta R_{3\varphi}$ (b) as a function of external field measured at 50 K, 200 K and 300 K, respectively. Solid-lines are fittings using Eq. (4) (for $\Delta R_\varphi$) and Eq. (5) (for $\Delta R_{3\varphi}$). (c) Temperature dependence of $A$, $B$, and $C$ from 50 K to 300 K. (d) Temperature-dependence of $\Delta R_{AMR}$, $\Delta R_1$ and $\Delta R_3$ at different $H_{ex}$.

### D. Pt thickness dependence of nonlinear magnetoresistance

The above results are for the NiFe(1.8)/Pt(2) bilayer sample. In order to confirm if the sign inversion of $\Delta R_{3\varphi}$ occurs only for samples with a specific Pt thickness, we fixed the NiFe thickness at 1.8 nm and varied the Pt thickness from 2 nm to 9 nm. All the samples were configured in the same form of Wheatstone bridge with four ellipsoidal elements. To facilitate comparison of samples with different Pt thicknesses, during the measurements, the current density in the Pt layer was fixed at $5.5 \times 10^5$ A/cm$^2$ (rms value) for all the samples. Again, it was found that, except for the low-field range, $\Delta R(\varphi)$ can be decomposed as $\Delta R(\varphi) = \Delta R_\varphi \sin \varphi + \Delta R_{3\varphi} \sin 3\varphi$, where $\varphi$ is the angle of in-plane field with respect to the current direction. Figures 5(a) and 5(b) show the experimentally extracted $\Delta R_\varphi$ and $\Delta R_{3\varphi}$ values (symbols), respectively, at different Pt thicknesses as a function of the external field. Solid-lines are fittings using Eq. (4) for $\Delta R_\varphi$ and Eq. (5) for $\Delta R_{3\varphi}$. As is with the case of NiFe(1.8)/Pt(2), $\Delta R_\varphi$ decreases monotonically with increasing the external field for all the samples with different Pt thicknesses. Despite the gradual decrease of $\Delta R_\varphi$ with the Pt thickness due to mainly the current shunting effect, its sign remains positive for all the samples. On the other hand, sign reversal of $\Delta R_{3\varphi}$ also occurs for all the samples even though the absolute value becomes very small for samples with a Pt thickness of 7 nm and 9 nm. We also change the NiFe thickness to 3 nm and 5 nm and keep Pt thickness at 2 nm. The overall trend remains the same. These results demonstrate that the model described by Eq. (1) is applicable to all samples regardless of the Pt and NiFe thicknesses.

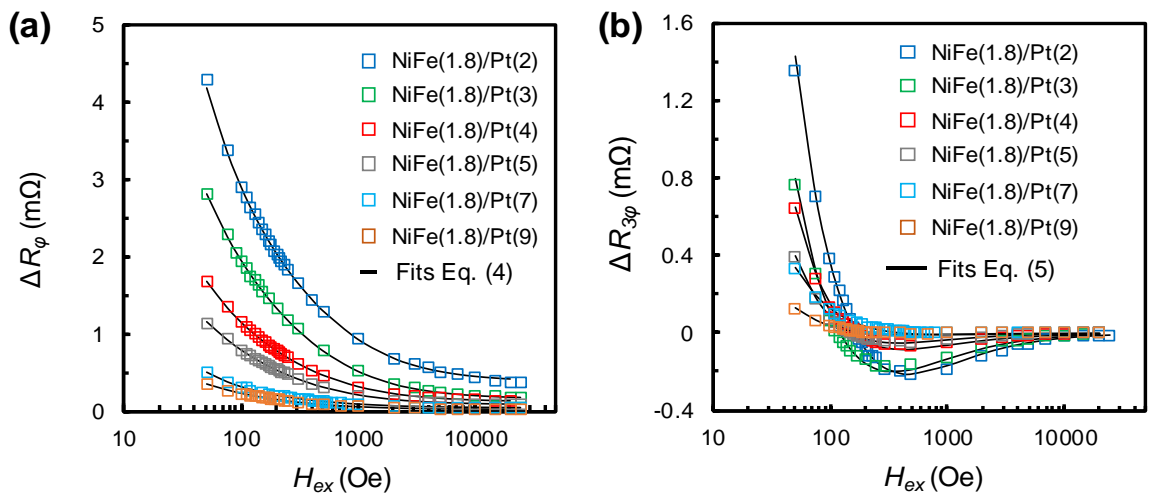

FIG. 5. (a) Experimentally extracted $\Delta R_\varphi$ (symbols) at different Pt thicknesses as a function of the external field. Solid-lines are fittings using Eq. (4). (b) Same as (a) but for $\Delta R_{3\varphi}$.

### E. Comparison with NiFe/Ta bilayers

In the aforementioned discussion, we have assumed that both the SOT and MMR are originated from the SHE effect in Pt. Hence it is expected that the fitting parameters, *A, B* and *C*, should reverse the sign when Pt is replaced by another metal with opposite spin Hall angle, such as Ta. To test this hypothesis, we fabricated a NiFe(1.8)/Ta(3) sample with the same Wheatstone bridge structure as that of the NiFe/Pt sample and performed the same angle dependence of 2$^{nd}$ harmonic MR measurement. The results of $\Delta R_\varphi$ obtained at external field strength of 500 Oe and 2T are shown in Fig. 6(a), together with the results of NiFe(1.8)/Pt(2). As can be seen, both samples exhibit similar angular dependence, but their signs are opposite with each other. When $H_{ex}$ increases to 2T, the bridge output of both samples contains only a $\sin\varphi$ component, which is mainly due to the spin-dependent USMR as discussed earlier. The thermoelectric effect is negligible in the present samples, as the bridge output for a NiFe(1.8)/Cu(3)/Ta(1) Wheatstone bridge at 2 T is almost zero (see Supplementary Material). This is also manifested in the fact that $\Delta R_\varphi$ at 2 T exhibits an opposite polarity for the NiFe/Ta and NiFe/Pt samples, which is consistent with the USMR scenario [17]. Figure 6(b) shows the decomposed $\sin 3\varphi$ component at 500 Oe. It shows clearly that the $\sin 3\varphi$ component also shows an opposite sign for the two types of samples. To quantify MMR in NiFe/Ta bilayers, we extract the nonlinear resistance components $\Delta R_\varphi$ and $\Delta R_{3\varphi}$ from the angle dependent 2$^{nd}$ harmonic MR measurement at external fields from 50 Oe to 2 T. The results are shown in Fig. 6(c) and Fig. 6(d), respectively, as a function of $H_{ex}$ for NiFe(1.8)/Pt(2) (square) and NiFe(1.8)/Ta(3) (triangle). The solid-lines are fittings for $\Delta R_\varphi$ using Eq. (4) and $\Delta R_{3\varphi}$ using Eq. (5). The parameters used for the fittings are $A = 130$ m$\Omega\cdot$Oe, $B = 637$ m$\Omega\cdot$Oe and $C = -433$ m$\Omega\cdot$Oe for NiFe(1.8)/Pt(2), and $A = -2.860$ m$\Omega\cdot$Oe, $B = -334.3$ m$\Omega\cdot$Oe and $C = 180$ m$\Omega\cdot$Oe for NiFe(1.8)/Ta(3). The signs are clearly opposite for the two samples. It is interesting to note that the absolute values of *B* and *C* for NiFe(1.8)/Ta(3) are nearly half of those of NiFe(1.8)/Pt(2); in contrast, *A* for NiFe(1.8)/Ta(3) is significantly smaller than that of NiFe(1.8)/Ta(3). The much smaller *A* in NiFe/Ta may be explained by the fact that the SOT effective field in NiFe/Ta is opposite to the Oersted field, whereas in NiFe/Pt, they are in the same direction. The partial cancellation of SOT effective field by the Oersted field may result in a smaller *A* as it is proportional to the net effective field. On the other hand, the MMR is not affected by the magnitude of the net effective field. These results provide additional evidence that the sign reversal of $\Delta R_{3\varphi}$ induced by external field is due to the competition between AMR/SMR and MMR.

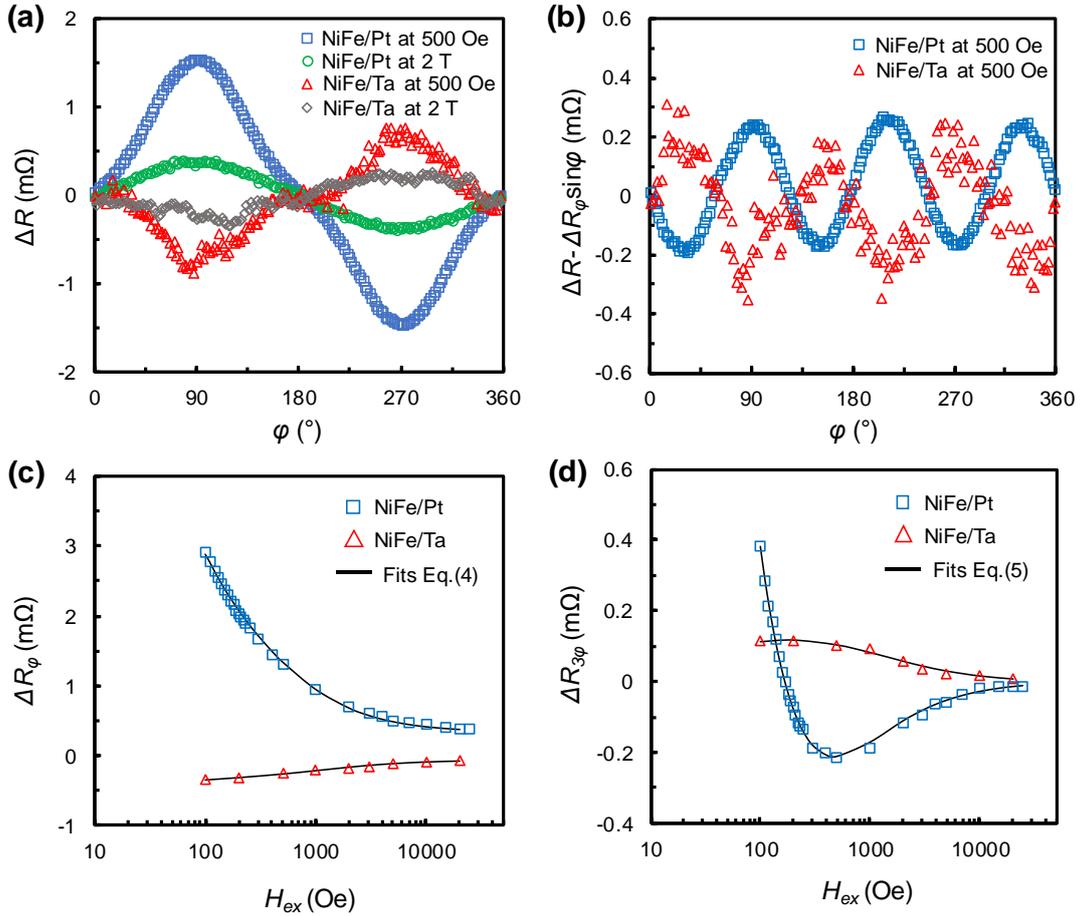

FIG. 6. (a) $\Delta R$ obtained at an applied field of 500 Oe and 2T for both NiFe(1.8)/Pt(2) and NiFe(1.8)/Ta(3) samples. (b) The $\sin 3\varphi$ component decomposed from the $\Delta R$ curves in (a) by subtracting out the $\sin \varphi$ component. (c,d) Nonlinear resistance components $\Delta R_\varphi$ (c) and $\Delta R_{3\varphi}$ (d) extracted by fitting the bridge output $\Delta R$ as a function of $H_{ex}$ for the NiFe(1.8)/Pt(2)

(square) and NiFe(1.8)/Ta(3) (triangle) samples. The solid-lines are fittings using Eq. (4) (for $\Delta R_\varphi$) and Eq. (5) (for $\Delta R_{3\varphi}$).

### F. Discussion

The above results all point to the fact that the 2$^{nd}$ order MR contains the contributions from AMR/SMR, SD-USMR, and MMR, which can be separated by the field strength and angle dependence data. Although the experimental data can be accounted for reasonably well using the model described by Eq. (1), first principles studies are required to unveil the true origin of the $\sin\varphi$ and $\sin 3\varphi$ terms of the MMR. As we demonstrated in this work and also our previous work [25], the bridge method is uniquely suited for measuring MR which is an odd function of the current. Due to the low noise in the signal, there is no need to perform any post-measurement processing of the data, which helps to increase the rigor of data analysis. We expect that it will become a powerful technique for characterizing spin texture of various types of materials.

## V. CONCLUSIONS

In conclusion, we have conducted a systematic angular dependence study of MR in NiFe/Pt bilayers at variable temperature and field and successfully disentangled MMR from the AMR and SMR. It is found that the angular-dependence of MMR contains two terms proportional to $\sin\varphi_m$ and $\sin 3\varphi_m$, respectively, and both terms scale as $(H_{ex} + H_m)^{-1}$. On the other hand, the AMR and SMR scale with the external field only as $1/H_{ex}$. The competition between MMR and AMR/SMR leads to a sign change of the non-linear magnetoresistance at a specific magnetic field, which was not reported previously. Furthermore, the measurement of NiFe/Ta bilayers further confirms the SHE origin of MMR in FM/HM bilayers. Our results provide further insights into MMR in FM/HM bilayers and demonstrate the importance of disentangling different types of MRs when characterizing charge-spin interconversion in FM/HM bilayers.

## ACKNOWLEDGEMENTS

Y.H.W. would like to acknowledge support by the Ministry of Education, Singapore under its Tier 2 Grant (Grant No. MOE2017-T2-2-011). Y.H.W. is a member of the Singapore Spintronics Consortium (SG-SPIN).